\documentclass[sn-aps]{sn-jnl}

\jyear{2021}%

\theoremstyle{thmstyleone}%

\theoremstyle{thmstyletwo}%

\theoremstyle{thmstylethree}%

\usepackage{amsmath}
\usepackage{xcolor}
\usepackage{array}
\usepackage{comment}

\begin{document}

\title{Observing short-range correlations in nuclei through $\rho^0$ photo-production}

\author*[1]{\fnm{Phoebe} \sur{Sharp}}\email{psharp15@gwu.edu}

\author[1]{\fnm{Axel} \sur{Schmidt}}\email{axelschmidt@gwu.edu}

\affil*[1]{\orgdiv{Department of Physics}, \orgname{The George Washington University}, \orgaddress{\street{725 21st St. NW}, \city{Washington},  \state{DC} \postcode{20052}, \country{USA}}}

\abstract{Short-range correlations (SRCs) are a universal feature of nuclear structure. A wide range of measurements, primarily using electron scattering, have revealed SRC properties, such as their abundance in different nuclei, as well as the strong preference for proton-neutron pairing over proton-proton or neutron-neutron pairing. Despite the inherent complexity of many-body systems, a number of the salient features of electron scattering measurements are described by a simple, factorized theory called Generalized Contact Formalism. A key element of this theory, the factorization of the interaction with a hard probe, has yet to be tested. An experiment conducted at Jefferson Lab in 2021 collected data from scattering a tagged photon beam, with an energy up to 10 GeV, from several nuclear targets, measuring final state particles in the large-acceptance GlueX spectrometer. In this paper, we propose a test of probe factorization by measuring cross section ratios sensitive to proton-proton pair prevalence and relative SRC abundances in $^4$He and $^{12}$C. We present GCF predictions of the observables and make projections of the expected precision the experiment can achieve.} 

\maketitle

\section{Introduction}\label{sec1}

Short-range correlations (SRCs) between nucleons within nuclei appear to be a universal feature of nuclear structure~\cite{Feldmeier:2011qy,Cruz-Torres:2019fum}, appearing in nuclei ranging in size from the deuteron~\cite{HallC:2020kdm} all the way to gold~\cite{Fomin:2011ng} and lead~\cite{Hen:2014nza}. 
Several calculations have shown that short-range correlations even influence the structure of infinite nuclear matter at high-density, indicating that they may affect the properties of neutron stars, such as size and cooling rates~\cite{Frankfurt:2008zv}. 
Unlike neutron star matter, SRCs have been studied in the lab, primarily through high-energy electron scattering. 
Electron scattering measurements have consistently shown that the density of SRCs in medium- and heavier-mass nuclei is 4--5 times higher than in deuterium~\cite{Fomin:2011ng,CLAS:2003eih,CLAS:2019vsb}, that nucleons in SRCs have high relative momentum, compared to the nuclear Fermi momentum~\cite{CLAS:2003jrg,HallC:2020kdm}, while also having a comparatively small center-of-mass momentum~\cite{CLAS:2018qpc}, and finally that SRCs form predominantly from spin-1 neutron-proton (np) pairs~\cite{Subedi:2008zz, JeffersonLabHallA:2007lly, LabHallA:2014wqo, CLAS:2018xvc, Li:2022fhh}.
This latter property is known as $np$-dominance. There is evidence that the prevalence of $np$ pairs changes with relative momentum, which may reveal the isospin dependence of the effective short-range nucleon-nucleon interaction~\cite{CLAS:2020mom,CLAS:2020rue}. 

Recently, the theoretical tool of Generalized Contact Formalism (GCF) has provided insights into SRCs as well as the interpretation of electron scattering experiments that seek to isolate them. 
Generalized Contact Formalism is a scale-separated theory that assumes a factorization of the structure of a correlated pair of nucleons from the rest of the nucleus~\cite{PhysRevC.92.054311,WEISS2018211}. This factorization works both in position space, where correlated nucleons are much closer to each other than any other nucleons, or in momentum space, where correlated nucleons tend to have higher relative momenta than the typical nuclear Fermi momentum~\cite{Cruz-Torres:2019fum}.
GCF has been used to calculate the high-momentum region of nuclear spectral functions~\cite{WEISS2019242} as well as the fully differential cross sections for the break up of SRCs in high-energy (multi-GeV) electron scattering~\cite{CLAS:2020mom}. 
GCF performs surprisingly well at describing the kinematic distributions of electron scattering data, in the limits where the momentum transfer is high, the missing momentum is high, and the missing momentum is oriented anti-parallel to the momentum transfer, so called ``anti-parallel kinematics'' \cite{CLAS:2020mom,CLAS:2020rue,Pybus:2020itv}.

One feature of Generalized Contact Formalism is a factorization of the hard scattering reaction between the probe and the struck nucleon from the nuclear physics that governs the distributions of nucleons within the nucleus. This implies that GCF should equally describe the results of high-energy electron scattering with those from other probes, provided the reaction is energetic enough to justify the factorization. GCF appears to match the results of high-energy proton-Carbon scattering~\cite{Patsyuk:2021fju}, but the probe independence prediction of GCF is otherwise untested.

In 2021, an experiment was conducted at Jefferson Lab to explore short-range correlations using high-energy photo-production~\cite{GlueX:2020dvv}. A tagged photon beam covering an energy range of approximately 6--10~GeV was scattered from deuterium, helium, and carbon targets with one of the goals being to find signatures of short-range correlations in a range of photo-production reactions. Photo-production offers an interesting complement to electron scattering. The kinematics of the reaction favor interactions with nucleons with initial momentum parallel to incoming photon direction, as opposed to the anti-parallel kinematics typically used in electron scattering. Furthermore, photon scattering through the $\gamma n \rightarrow \pi^- p$ or $\gamma n \rightarrow \rho^- p$ channels offers the possibility of probing neutrons without having to detect them in the final state. At the time of this writing, the data from this experiment are under analysis. 

In this paper, we use GCF to calculate predictions for two observables sensitive to properties of short range correlations that have been studied in electron scattering. We specifically consider the $\gamma p \rightarrow \rho^0 p$ reaction, one of the most straight-forward photo-production reactions that the recent Jefferson Lab experiment has measured. First, we consider the relative abundance of short-range correlations in deuterium, helium, and carbon by comparing the relative rates of $\rho^0$ production in kinematics where the struck nucleon likely came from an SRC. Second, we consider the coincident detection of a recoiling spectator proton. The prevalence of such protons can provide information about the relative abundance of $pp$ and $pn$ SRCs. The Jefferson Lab experiment will have adequate statistics to make impactful statements from both observables. 

\section{The Short Range Correlations / Color Transparency Experiment}\label{sec:exp}

The Short-Range Correlations / Color Transparency (SRC/CT) experiment (Jefferson Lab experiment E12-19-003) took place in Experimental Hall D of Jefferson Lab in Newport News, Virginia, during the fall of 2021. 
The Hall D photon beam was scattered from liquid deuterium, liquid helium, and a multi-foil carbon target, with data being collected by the GlueX spectrometer~\cite{GlueX:2020idb}, a large acceptance spectrometer with a 2~T solenoid magnet. The trigger for this experiment was designed to have minimum bias for the wide range of photoproduction reactions being considered. The trigger required a minimum energy deposition in the Barrel Calorimeter (BCal) and Forward Calorimeter (FCal) in coincidence with a charged particle hit in the Start Counter (SC) surrounding the target. The total photon flux was monitored using a Pair Spectrometer (PS), with its own trigger, positioned upstream of the target. 
A break down of the run time, triggers, and PS triggers collected during the experiment is given in Table~\ref{table:triggers}.

\begin{table}
\begin{center}
\caption{Production data collected in the  SRC/CT Experiment}
\begin{tabular}{ c c c c }
\hline
\hline
Target & Duration [days] & Triggers [billions] & PS Triggers [billions]\\
\hline
Deuterium & 4 & 16.4 & 0.82 \\
Helium & 10 & 29.5 & 2.05\\ 
Carbon & 14 & 46.7 & 3.20\\
\hline
\end{tabular}
\label{table:triggers}
\end{center}
\end{table}

The SRC/CT Experiment had two primary goals: exploring short-range correlations using photo-production reactions as well as searching for evidence of color transparency in photo-production. Both can be studied in a wide range of photo-production reactions, a partial list of which can be seen in Table~\ref{table:channels}. In this paper, we consider only the $\gamma p \rightarrow \rho^0 p$ reaction, as it has a comparatively high cross section and because it is straight-forward to identify through the $\rho^0$ meson's decay to $\pi^+\pi^-$. In addition to those primary goals, data from the experiment can also be used to search for Axion-Like Particles (ALPs)~\cite{Pybus:2023yex}, search for evidence of in-medium branching ratio modification, and probe gluonic form factors~\cite{Pybus:2024ifi}. 

\begin{table}
\caption{Reactions that can be studied in the SRC/CT Experiment}
\begin{center}
\begin{tabular}{ c c }
\hline
\hline
 Proton Reactions & Neutron Reactions \\ 
 \hline
 $\gamma + p \rightarrow \pi^0 + p$ & $\gamma + n \rightarrow \pi^- + p $\\  
 $\gamma + p \rightarrow \pi^- + \Delta^{++}$ &$ \gamma + n \rightarrow \pi^- + \Delta^{+} $\\    
  $\gamma + p \rightarrow \rho^0 + p$ & $\gamma + n \rightarrow \rho^- + p $\\    
  $\gamma + p \rightarrow K^+ + \Lambda^{0} $&$ \gamma + n \rightarrow K^0 + \Lambda^{0}$ \\    
   $\gamma + p \rightarrow K^+ + \Sigma^{0} $& $\gamma + n \rightarrow K^0 + \Sigma^{0}$ \\    
    $\gamma + p \rightarrow \omega + p$ &$\gamma + n \rightarrow K^+ + \Sigma^{-}$ \\
    $ \gamma + p \rightarrow \phi + p$ & \\
\hline
\end{tabular}
\label{table:channels}
\end{center}
\end{table}

\section{Generalized Contact Formalism}
\label{sec:sim}

\begin{figure}
    \centering
    \includegraphics{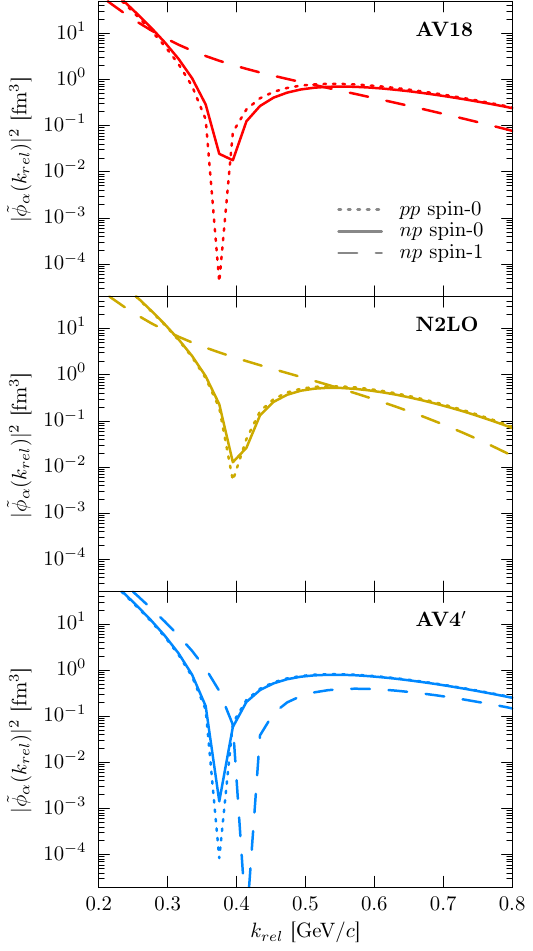}
    \caption{\label{fig:phiSq}
    The relative momentum distribution for nucleons in a short-range correlated pair   
    used in the GCF calculations of this work. These distributions are determined from
    the zero-energy solution of the Schr\"{o}dinger Equation for a given $NN$-potential
    model. While the Argonne AV18~\cite{PhysRevC.51.38}, and $\chi$EFT (N2LO)~\cite{Gezerlis:2013ipa}
    potentials are realistic, the Argonne AV4$'$ potential~\cite{Wiringa:2002ja} is a simplified potential
    with no tensor term. See text for details.
    }   
\end{figure}

In this analysis, we make predictions for $\rho^0$ photo-production observables based on the theoretical model of Generalized Contact Formalism. 
Generalized Contact Formalism (GCF) is a factorized description of short-range correlations within nuclei~\cite{PhysRevC.92.054311,WEISS2018211,WEISS2019242}.
GCF relies on the separation of scales between the correlated two-nucleon system and the remaining nucleons in the nucleus. This separation occurs both in position space (the correlated nucleons have much closer proximity to each other) and momentum space (the relative momentum of correlated nucleons can be larger than the typical Fermi momentum). Scale separation is used to justify a factorization the nuclear wave function:
\begin{equation}
\Psi  \xrightarrow[]{r_{ij} \rightarrow 0} 
\sum\limits_\alpha \phi_{ij}^\alpha(\vec{r}_{ij})
A_{ij}^\alpha(\vec{R}_{ij},\vec{r}_{k\neq i,j}),
       \label{eq:wavefun}
\end{equation}
where $i$ and $j$ label the correlated nucleons, $r_{ij}$ is their separation, $R_{ij}$ is the position of their center of mass, $\alpha$ labels their relative quantum numbers (i.e., spin, isospin), $A_{ij}$ describes the wave function of the rest of the nucleons $k\neq i,j$, and  ${\phi}(r_{ij})$ is an asymptotic two-body wave function that is independent of nucleus. This ansatz can be used to formulate one- and two-body nucleon momentum densities~\cite{PhysRevC.92.054311,WEISS2018211}, nuclear spectral functions~\cite{WEISS2019242}, and hard scattering cross sections~\cite{CLAS:2020mom,Pybus:2020itv} for nucleons in the short-range correlated regime. GCF cross sections take a factorized form:
\begin{equation}
    d \sigma \approx \sigma_\text{probe}  \sum\limits_{\alpha} C_{\alpha}
    \cdot P_{\alpha}(\vec{k}_{cm}) \cdot \lvert \tilde{\phi}_{\alpha}(\vec{k}_{rel})\rvert ^2,
    \label{eq:GCF}
\end{equation}
where $\sigma_\text{probe}$ describes the cross section for the hard interaction between a probe and a single nucleon, $P_\alpha(\vec{k}_{cm})$ describes the distribution of the center-of mass momentum $\vec{k}_{cm}$ of the pair, $\lvert \tilde{\phi}_{\alpha}(\vec{k}_{rel})\rvert ^2$ describes the nucleus-independent relative momentum ($\vec{k}_{rel}$) distribution of the correlated nucleons, and $C_\alpha$ is the ``Contact,'' effectively the abundance of SRC pairs with quantum numbers $\alpha$ in the nucleus. (More formally the contact is defined by: $C=16\pi^2 \sum_{ij}\langle A_{ij} \lvert A_{ij}\rangle $.) 
Each of these components in the equation must be specified from some external information. 

\begin{table}[]
    \caption{Parameter values used with Eq.~\ref{eq:SLAC}}
    \centering
    \begin{tabular}{c l}
    \hline
    \hline
    Parameter & Value\\
    \hline
   A & 90057 nb\\
   B & 5.941 GeV$^{-2}$\\
   C & $1.097\cdot 10^{8}$ nb$\cdot$GeV$^{14}$ \\ 
   D & 3.834 \\
   E & 1.795 \\
    \hline
    \end{tabular}
    \label{tab:SLAC}
\end{table}

We have produced GCF predictions using Monte Carlo simulation with an event generator whose cross section takes the form of Eq.~\ref{eq:GCF}. For the single-nucleon $\gamma p \rightarrow \rho^0 p$ cross section, we assume the parametrization
\begin{equation}
    \sigma_{\gamma p \rightarrow \rho^0 p} = A \exp{B t}+ C s^{-7} ( 1.2 - \cos{\theta_{cm}})^{-D}(1.05 +  \cos{\theta_{cm}} ) ^{-E}
    \label{eq:SLAC}
\end{equation}
with Mandelstam variables $s$ and $t$, with $\cos\theta_{cm}$ as the $\rho$ scattering angle in the center-of-mass frame, and the coefficients $A$, $B$, $C$, $D$, and $E$ determined from fits to data with $E_\gamma=4$ and $6$~GeV collected at SLAC~\cite{Anderson:1976ph} (values shown in Table~\ref{tab:SLAC}).

Contacts for deuterium, helium, and carbon were taken from Ref.~\cite{Cruz-Torres:2019fum}, which were determined from fits of variational Monte Carlo calculations of two-body momentum densities~\cite{Wiringa:2013ala,Piarulli:2022ulk}.

The pair center-of-mass motion distribution, $P_\alpha(\vec{k}_{cm})$ was assumed to be $\alpha$-independent and Gaussian, with $\sigma=100$~MeV$/c$ for helium and $\sigma=150$~MeV$/c$ for carbon based on the experimental measurements of Refs.~\cite{LabHallA:2014wqo,CLAS:2018qpc}. For deuterium, there is no center-of-mass momentum of the pair. 

Finally, the universal two-body momentum distributions, $\lvert \tilde{\phi}(\vec{k}_{rel})\rvert ^2$, were determined from zero-energy solutions to the Schr\"{o}dinger equation (described in \cite{WEISS2018211}), under an assumed nucleon-nucleon potential. In this work, we primarily consider the AV18 potential~\cite{PhysRevC.51.38}, a realistic phenomenological potential, as well as a potential derived from chiral effective field theory ($\chi$EFT), at the N2LO level, developed in Ref.~\cite{Gezerlis:2013ipa}. We also consider AV4$'$, which is a highly simplified phenomenological potential that, significantly, contains no tensor term~\cite{Wiringa:2002ja}. Differences in behavior between AV18 and AV4$'$ can highlight the impact of the tensor force. The universal two-body momentum distributions for the three interactions are shown in Fig.~\ref{fig:phiSq}.

To perform the calculations, we generated $A(\gamma,\rho^0 pN)$ events according to the GCF cross section and then applied several event selection criteria to classify events. We assumed a photon beam energy of $8$~GeV, which is approximately the coherent peak energy in the SRC/CT Experiment.
Our event selection uses missing momentum, $p_\text{miss}$, as a proxy for the initial momentum of the struck nucleon prior to the interaction. We define missing momentum as:
\begin{equation}
    p_\text{miss} \equiv \left\lvert \vec{p}_p + \vec{p}_{\rho^0} - \vec{p}_\gamma \right\rvert
\end{equation}
where $\vec{p}_p$ is the final outgoing momentum of the struck proton (the event generator specifies which nucleon was struck), $\vec{p}_{\rho^0}$ is the outgoing momentum of the $\rho^0$ meson, and $\vec{p}_\gamma$ is the photon beam momentum. 

First, we selected a sample of $A(\gamma,\rho^0 p)$ events, using the following criteria:
\begin{itemize}
    \item $|t|>1.5$ GeV$^2$ and $|u|>1.5$ GeV$^2$, where $t$ and $u$ are the Mandelstam variables. These criteria enforce that the reaction is sufficiently hard to justify the factorization implicit in GCF.
    \item $p_{\text{miss}}>0.350$ GeV. A lower bound on $p_{\text{miss}}$ ensures that the struck nucleon came from a momentum state above the nuclear Fermi momentum and is thus likely to be in a short-range correlated pair.
    \item  $p_p > 1$ GeV, in order to avoid confusion between the struck proton (high-momentum) and any emitted spectator protons (low-momentum).  
\end{itemize}

From the events that satisfy the above criteria, we isolate a subset of $A(\gamma,\rho^0 pp)$ events that meet the following additional event selection criteria:
\begin{itemize}
    \item The event contains a second proton with momentum less than $0.8$~GeV$/c$. In GCF, this proton is assumed to a be a spectator to the reaction, recoiling from the nucleus once its correlated partner has been removed due to the hard interaction with the probe. We label this proton's momentum as $\vec{p}_\text{recoil}$.
    \item $p_\text{recoil} > 0.3$ $GeV/c$. This requirement ensures the proton has enough momentum to exceed the GlueX detection threshold.
\end{itemize}

\section{Projected Statistical Uncertainty}
\label{sec:data}

In order to estimate the statistical uncertainty achievable in the SRC/CT Experiment, we performed a preliminary analysis attempting to isolate the yields of $A(\gamma, \rho^0 p )$ and $A(\gamma, \rho^0 p p )$ events in the total data set. After being calibrated, the data were passed through the standard GlueX reconstruction software and filtered for a final state consisting of $\pi^+\pi^-$ and one or two protons. The particle momentum vectors in each event were re-fit to require a common vertex within the target volume. This re-fit slightly enhances the momentum resolution of the reconstruction (approximately a 5--7\% improvement in $p_\text{miss}$ resolution, depending on kinematics).

The GlueX experiment uses a tagged photon beam. Each event is correlated with an electron detected in the tagger detector. 
A significant fraction of events come from random coincidences between the tagger and GlueX. We adopted a procedure used
in GlueX to subtract this random coincidence background. We determined the rate of so-called ``off-time'' events, i.e., events in which the electron detection in the tagger was either too early or too late to be associated with the GlueX trigger. 
Off-time events were proportionally subtracted from any histograms we produced.

One of the challenges in isolating the $\rho$-photoproduction reaction from other background processes is particle identification (PID). 
The GlueX spectrometer is not able to cleanly separate high-momentum protons and positive pions, starting at about 1~GeV$/c$. 
When studying exclusive reactions from a hydrogen target, momentum and energy conservation conditions can be deployed to resolve ambiguities in PID. This technique is not possible when studying a nuclear target, where the undetected residual nucleus can act as a reservoir of energy and momentum.

This leads to situations where the proton and $\pi^+$ in the event are inadvertently switched. It can also allow a background process, the simultaneous production of both a $\rho^0$ and $\pi^+$, to appear as a $\rho^0$ and proton, i.e., the signal. To make an estimate of statistical precision, we have taken a rudimentary approach to removing these sources of background. The methods will be refined, and their consequences for systematic uncertainty studied, in the further analyses of these data. 

\begin{figure}
    \centering
    \includegraphics[width=5.55cm]{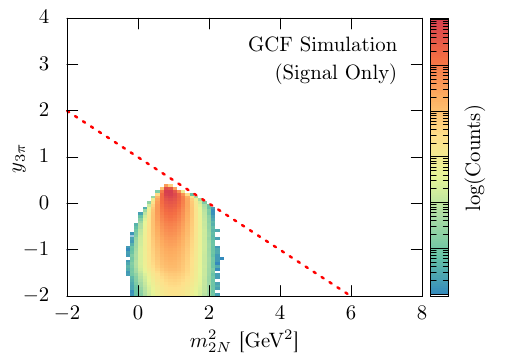}
    \includegraphics[width=5.55cm]{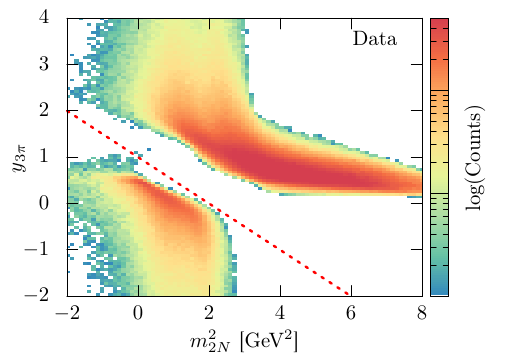}
    \caption{Left: Distribution of events according to our signal-only GCF simulation for the reaction
    $^{12}C(\gamma, \pi^+ \pi^- p)$ using two-body wave functions, $\tilde{\phi}_\alpha$, produced using the AV18 potential.
    The criteria listed in Section~\ref{sec:sim} have been applied.
    Right: $^{12}C(\gamma, \pi^+ \pi^- p) X$ data collected during the experiment subject to the beam energy,
        vertex position, proton angle, and missing momentum criteria listed in Section~\ref{sec:data}. At this 
        stage in the analysis, only a $\lvert t\rvert >1$~GeV$^2$ restriction has been applied. Accidental background
        has been subtracted. The diagonal dashed line denotes the cut placed on the data. }
   \label{fig:mxy3pi}
\end{figure}

To reduced background from $\rho^0 \pi^+$ production appearing as $\rho^0 p$ signal, we developed several new variables that appear to provide clean separation. First, we use the four-momentum vector $p_{\pi^+ \rightarrow p}^\mu$ to represent the four-momentum vector: $\left(\sqrt{p_p^2 + m_\pi^2},\vec{p}_p \right)$. If the particle reconstructed as a proton in the event were actually a pion, $p_{\pi^+ \rightarrow p}^\mu$ would be its four-momentum vector, since the energy is inferred from the measured momentum. We next define $t_{3\pi}$:
\begin{equation}
    t_{3\pi} \equiv (p_\gamma^\mu - p_{\pi^+}^\mu - p_{\pi^-}^\mu- p_{\pi^+ \rightarrow p}^\mu)^2, 
\end{equation}
the Mandelstam-$t$ for the reaction $\gamma p \rightarrow \rho^0 \pi^+ n$. Then, we define $y_{3\pi}$:
\begin{equation}
    y_{3\pi} \equiv \frac{-t_{3\pi}}{ 2m_N\left(E_\gamma - E_{\pi^+} - E_{\pi^-} -\sqrt{p_p^2 + m_\pi^2}\right)}, 
\end{equation}
a ratio of four-momentum transfer to energy transfer in the background process. 
Here, $m_N$ is the nucleon mass. Finally, we define the two-nucleon missing mass squared,
\begin{equation}
M_{2N}^2 \equiv (p_\gamma^\mu - p_{\pi^+}^\mu - p_{\pi^-}^\mu - p_p^\mu)^2 - 2 m_N (E_\gamma - E_\pi^+ - E_\pi^- - E_p) + 4 m_N^2,
\end{equation}
which assumes that the reaction proceeded from a nucleon whose initial momentum was exactly balanced by one other nucleon. We find that, on their own, neither $y_{3\pi}$ nor $M_{2N}^2$ is especially effective at separating signal from background. However, the two variables in combination is highly effective. This can be seen in Fig.~\ref{fig:mxy3pi}, in which the distribution of signal events predicted by our GCF simulation (left) is compared to the experimental data (right). A two-dimensional cut removes background present in data, but absent from the signal distribution.

We applied the following event selection criteria to the reconstructed data set to select our $A(\gamma,\rho^0 p)$ events:
\begin{itemize}
\item 6 GeV $< E_{\gamma} < 10.8$~GeV: to limit our analysis to photons within the tagger acceptance.
\item One proton (referred to as the leading proton) with momentum, $p_p$, greater than 1~GeV$/c$.
\item The leading proton's angle, $\theta_p > 10^\circ$, since simulations indicate that protons going forward of $10^\circ$ is highly unlikely for our signal reaction.
\item $-3$ GeV $< E_{\pi^+} + E_{\pi^-} + E_{p} - E_\gamma - 2m_N < $ 3 GeV, a loose requirement on the energy balance between the initial
and final state, in order to help resolve signal from random coincidence background.
\item An event vertex position between $z=50$~cm and $z=80$~cm in the GlueX coordinate system, encompassing the position of the target, and removing background from scattering from other material in the path of the photon beam.
\item $M_{\pi^+ p} > 1.3$~GeV$/c^2$ to remove possible background from $\Delta^{++}$ decay.
\item $M_{\pi^- p} > 1.3$~GeV$/c^2$ to remove possible background from $\Delta^{0}$ decay.
\item $p_\text{miss} > 0.4$~GeV$/c$, where $p_\text{miss}$ is calculated as $\lvert \vec{p}_p + \vec{p}_{\pi^+} + \vec{p}_{\pi^-} - p_\gamma\rvert$, to select events in which the leading proton was likely to be part of an SRC pair. 
\item $y_{3\pi}<1 - 0.5 M_{2N}^2/$GeV$^2$ (shown in Fig.~\ref{fig:mxy3pi}), to reduce background from the combined production of $\rho^0 \pi^+$, with the $\pi^+$ mis-reconstructed as a proton.
\item $0.57~\text{GeV}/c^2 < M_{\pi^+\pi^-} < 0.97$~GeV$/c^2$, to select the $\rho^0$ mass region.
\end{itemize}

\begin{figure}
    \centering
    \includegraphics[width=8cm]{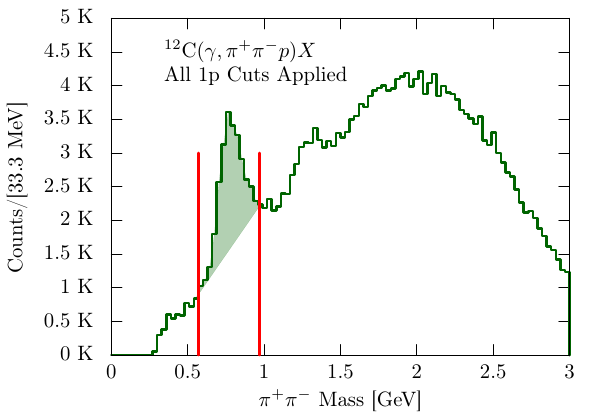}
    \caption{Invariant mass of the $\pi^+\pi^-$ pair for $^{12}C(\gamma, \pi^+ \pi^- p) X$ events measured in the experiment, subject
    to the $A(\gamma,\rho^0 p)$ selection criteria listed in Section~\ref{sec:data}. The red lines indicate our chosen mass range for
    the $\rho^0$ meson. }    
   \label{fig:rho0mass}
\end{figure}

\begin{figure}
    \centering
    \includegraphics[width=8cm]{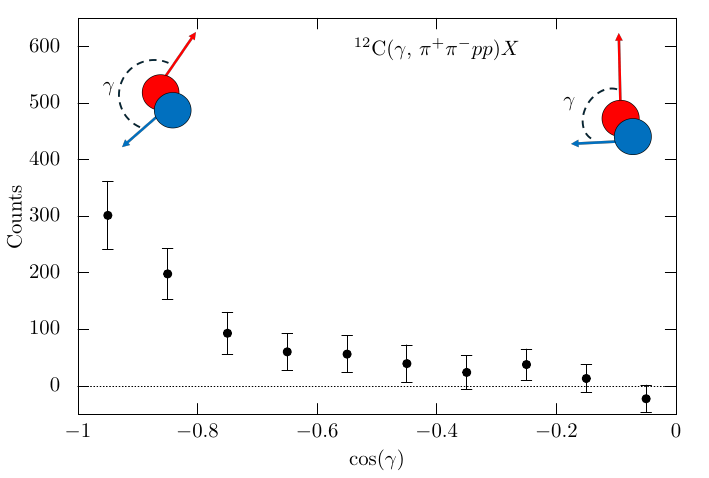}
    \caption{ Distribution of $\cos(\gamma)$, defined as the cosine of the angle between $\vec{p}_\text{miss}$ and $\vec{p}_\text{recoil}$
    for sideband-subtracted $^{12}$C$(\gamma, \pi^+ \pi^- pp)$ events. $\cos(\gamma)=-1$, where the majority of events lie,
    corresponds to two nucleons initially in a back-to-back configuration. Similar enhancements near $\cos(\gamma)=-1$ have been 
    observed as a signature of SRCs in quasi-elastic electron scattering experiments. }
   \label{fig:cosgamma}
\end{figure}

From these events, we select a subset for our $A(\gamma, \rho^0 pp)$ sample by further requiring:
\begin{itemize}
\item One other proton with momentum in the range: $0.3 \text{ GeV}/c < p_\text{recoil} < 0.8$ GeV$/c$ 
\end{itemize}

We estimated the yield of $\rho^0$ events above non-resonant $\pi^+\pi^-$ background by using the side-bands of the $\pi^+\pi^-$ invariant mass distribution, as shown in Fig.~\ref{fig:rho0mass}. We estimated the statistical uncertainty on that yield, including the statistical uncertainty of the background.

We were able to test if our $A(\gamma, \rho^0 pp)$ is predominantly populated by events coming from the break-up of SRC pairs by looking at the distribution of $\cos(\gamma)$, the cosine of the angle between $\vec{p}_\text{miss}$ and $\vec{p}_\text{recoil}$. In the limit of zero final-state interactions, $\gamma$ describes the orientation of the momentum vectors of the two nucleons prior to the reaction. SRCs are predicted to have nearly anti-parallel momenta ($\cos(\gamma) =-1$), and this has been observed in electron scattering measurements~\cite{CLAS:2003jrg,JeffersonLabHallA:2007lly,LabHallA:2014wqo}. The distribution of events (after the subtraction of accidental and non-resonant backgrounds) is shown in Fig.~\ref{fig:cosgamma}. The enhancement of events for anti-parallel angles is evidence that our preliminary analysis is selecting an event sample predominantly containing $pp$-SRCs, and that our yields can be used to make an informative estimate of the statistical precision that a full analysis could achieve. 

\section{Results}
\label{sec:results}

We consider two observables that can be used to verify properties of short-range correlations determined in electron scattering. Both are cross section ratios, which are more robust than cross sections to systematic effects such as those associated with luminosity and detector efficiency. The first observable is the ratio of 
two proton events to single proton events (with or without a second proton), i.e.,
\begin{equation}
    \frac{\sigma[C(\gamma, \rho^0 p p )]}{\sigma[C(\gamma, \rho^0 p )]}.
\end{equation}
This observable is sensitive to the abundance of $pp$-SRCs relative to $np$-SRCs, but avoids the technical challenge of neutron detection, for which GlueX has a far lower efficiency compared to charged particles and photons. Events with a reaction on a proton in an $np$-SRC pair can still end up in the $(\gamma, \rho^0 p )$ event sample. 
The exact ratio value is highly sensitive, however, to the detection efficiency for spectator protons in the 0.3--0.8 GeV$/c$ range, and this must be studied thoroughly to estimate systematic uncertainty. The approach has been used in prior electron-scattering experiments~\cite{JeffersonLabHallA:2007lly, CLAS:2020mom}, with systematic uncertainties on the order of 5--10\%. The azimuthal symmetry of the GlueX detector should make the recoil proton acceptance more uniform than in, for example, the CLAS Spectrometer~\cite{CLAS:2020mom}, and thus the systematic uncertainties more favorable. 

\begin{figure}[htbp]
    \centering
    \includegraphics[width=10cm]{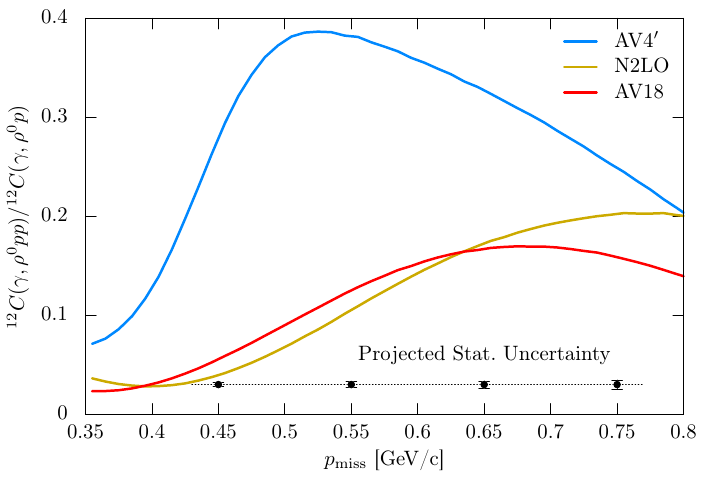}
    \caption{\label{fig:pp2p}
    GCF predictions for the cross section ratio $^{12}$C$(\gamma,\rho^0 pp)/^{12}$C$(\gamma,\rho^0 p)$ using three different $NN$ potential
    models as input, as well as the projected statistical uncertainty for this reaction anticipated for the SRC/CT Experiment. 
    No projection for systematic uncertainties of the experiment is shown.
    }
\end{figure}

The GCF predictions for this observable are shown in Fig.~\ref{fig:pp2p} for the three $NN$ interaction models used as input. Both the AV18 and $\chi$EFT-based N2LO potentials lead to predictions of a small $\sigma[C(\gamma, \rho^0 p p )]/\sigma[C(\gamma, \rho^0 p )]$ ratio, which increases slightly with increasing $p_{miss}$. The suppression of this ratio is due to the tensor part of the $NN$ interaction, leading to a suppression of $pp$-SRC pairs. This is confirmed by the prediction using the AV4$'$ potential, which has no tensor force, and consequently, predicts a much higher ratio.

Also shown in Fig.~\ref{fig:pp2p} are estimates of the projected statistical uncertainty achievable in the SRC/CT Experiment data set, in four bins of $p_\text{miss}$. The statistical uncertainties are easily small enough to distinguish the tensor-less and realistic potential models.
Provided that the systematic uncertainties can be kept at a similar level, $\rho^0$ photo-production could provide confirmation of $np$-dominance. 
Increasing the number of bins in the analysis would be preferable, though may not be possible, due to limited resolution on $p_\text{miss}$. Preliminary estimates indicate a $p_\text{miss}$ resolution on the order of 30--50~MeV$/c$, over the full range of kinematics. An optimization of the number of bins to balance the information gained against the systematic uncertainties from finite detector resolution should be studied further. 

\begin{figure}[htbp]
    \centering
    \includegraphics[width=10cm]{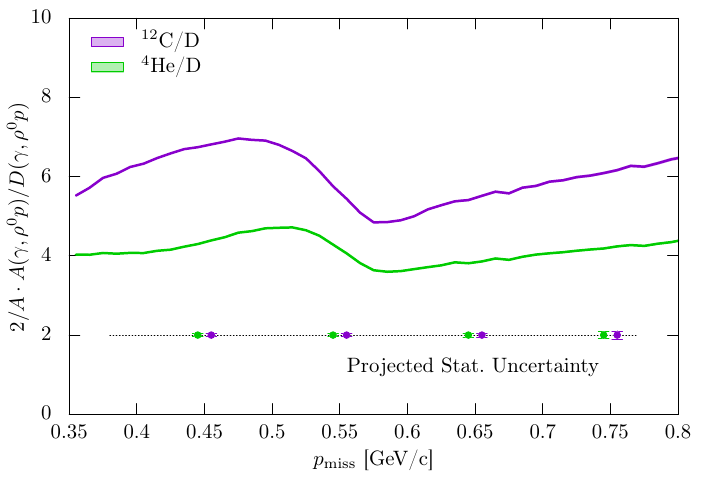}
    \caption{\label{fig:Aoverd}
    GCF predictions for the per-nucleon cross section ratio $(2/A)\cdot A(\gamma,\rho^0 p)/D(\gamma,\rho^0 p)$ 
    using the AV18 potential as input are shown for carbon-12 and helium-4, targets used in the SRC/CT Experiment.
    }
\end{figure}

The second observable is
\begin{equation}
    \frac{\sigma[A(\gamma, \rho^0 p )]}{\sigma[D(\gamma, \rho^0 p )]}
\end{equation}
where  $A$ represents either helium or carbon. This cross section ratio is sensitive to the relative abundance of SRC pairs in different nuclei, which has long been estimated using inclusive electron scattering~\cite{CLAS:2003eih,Fomin:2011ng,CLAS:2019vsb}. Electron scattering results show a plateau in the inclusive cross section ratio for Bjorken scaling variable $x_B > 1.4$. This plateau is evidence of a kinematic regime dominated by two-body (i.e., SRC-correlated) physics, rather than nucleus-dependent physics. Our GCF predictions for helium-4 and carbon-12 (using the AV18 potential) are shown in Fig.~\ref{fig:Aoverd}, along with estimates of the statistical uncertainty in a four-bin analysis. In contrast to the electron scattering results, GCF predicts some structure in this ratio. Depending on the systematic uncertainty and the eventual binning allowed by the experimental resolution in $p_\text{miss}$, this prediction of structure might be testable with data from the SRC/CT experiment. The largest sources of systematic uncertainty will likely come from background contamination, the background subtraction procedure, and experimental resolution.

\section{Conclusions}
\label{sec:conclusions}

In this paper, we lay out an approach for using $\rho^0$ photo-production from nuclei, measured in the SRC/CT Experiment, to validate properties of short-range correlations observed in electron scattering, and to test the limits of probe factorization assumed in Generalized Contact Formalism. The $\rho^0$ photo-production reaction is one of the most straight-forward reactions to consider, due to its fully charged final state, the easily identifiable decay of the $\rho^0$ into a $\pi^+\pi^-$ pair, and its high cross-section due to vector meson dominance. We make GCF predictions for two observables. Measuring the ratio of two-proton to one-proton events is a simple test of $np$-dominance. Ratios between nuclear targets, e.g., $^{12}$C$/D$ and $^{4}$He$/D$ offer a way to quantify the relative abundance of SRCs. Data analysis is underway and results are forthcoming. 

\bmhead{Acknowledgments}

This work was supported by the US Department of Energy Office of Science, Office of Nuclear Physics, under contract no. DE-SC0016583, the US Department of Energy Office of Science Graduate Student Research Program (2022--23), and the Jefferson Lab / Jefferson Science Associates Graduate Fellowship Program (2021--22). 

\bibliography{references}

\end{document}